\documentclass[12pt]{iopart}

\usepackage{graphicx}
\usepackage[colorlinks,citecolor=blue]{hyperref}

\begin{document}

\title{Topology optimization of 3D flow fields for flow batteries}

\author{Tiras Y. Lin, Sarah E. Baker, Eric B. Duoss, \& Victor A. Beck}

\address{Lawrence Livermore National Laboratory, Livermore, CA 94550, United States}
\ead{lin46@llnl.gov, beck33@llnl.gov}
\vspace{10pt}
\begin{indented}
\item[]February 2022
\end{indented}

\begin{abstract}
As power generated from renewables becomes more readily available, the need for power-efficient energy storage devices, such as redox flow batteries, becomes critical for successful integration of renewables into the electrical grid. 
An important component in a redox flow battery is the planar flow field,  which is usually composed of two-dimensional channels etched into a backing plate.
As reactant-laden electrolyte flows into the flow battery, the channels in the flow field distribute the fluid throughout the reactive porous electrode.
We utilize topology optimization to design flow fields with full three-dimensional geometry variation, i.e., 3D flow fields. 
Specifically, we focus on vanadium redox flow batteries and use the optimization algorithm to generate 3D flow fields evolved from standard interdigitated flow fields by minimizing the electrical and flow pressure power losses.
To understand how these designs improve performance, we analyze the polarization of the reactant concentration and exchange current within the electrode to highlight how the designed flow fields mitigate the presence of electrode dead zones.
While interdigitated flow fields can be heuristically engineered to yield high performance by tuning channel and land dimensions, such a process can be tedious; this work provides a framework for automating that design process. 
\end{abstract}

%
%
%
%
%

\section{Introduction}

The need for effective, large-scale energy storage continues to grow as the technology driving renewables continues to mature \cite{chu2012opportunities,chu2017path,Gur:2018ct}. 
Renewable sources, such as wind and solar, are inherently weather and time dependent. 
Unfortunately, this intermittent energy supply is not always aligned with user needs for electricity, necessitating the use of power plants to meet peak demand \cite{castillo2014grid,ela2011operating}. 
The mismatch between supply and demand hampers our ability to completely integrate renewables into the electrical grid. 
A stable, integrated electrical grid requires effective, large-scale energy storage -- that is, efficient conversion from electrical energy to potential (e.g. chemical) energy, and vice versa -- to buffer against this mismatch by storing and dispatching any excess generated electricity as necessary \cite{chen2009progress, Gur:2018ct}. 
Additionally, to be industrially relevant, these storage devices must operate with high power efficiency while remaining economically practical \cite{kear2012development}. 
Redox flow batteries inherently decouple energy storage from power production, allowing the former to scale independently by merely increasing the size of the external storage vessels.
As a result, redox flow batteries are an especially promising energy storage system to meet these needs \cite{weber2011redox,soloveichik2015flow,wang2013recent,perry2015advanced}.

A critical problem in designing large-scale electrochemical systems, and energy storage systems such as redox flow batteries in particular, is the distribution of reactants to the electrode. 
The electrode is a porous media with a large exposed surface area of electrochemically active material \cite{qiu20123}. 
Often, the electrode is made of a stochastic, graphitic or vitreous carbon material cohered into a bulk material such as a paper or felt \cite{weber2011redox}. 
Ideally, the electrolyte flow to the electrode will minimize excess power losses while appropriately distributing the reactant to promote reaction and mitigate the emergence of dead zones (i.e., low concentration regions with insufficient mass transfer) \cite{knudsen2015flow,ke2018rechargeable,nemani2017uncovering,ghimire2018comprehensive,wong2021direct}. 
This problem is usually addressed by using sophisticated fluid distribution systems both external and internal to the electrochemical cell stack. 
The internal fluid distribution system is known as a flow field \cite{ke2018rechargeable,arenas2019redox}, and significant effort has been devoted to understanding how the flow field architecture impacts device performance \cite{kumar2016effect,messaggi2018analysis,maurya2018effect,latha2014hydrodynamic,houser2016influence,darling2014influence,li2005review,knudsen2015flow}.

Two of the most prevalently used flow field architectures are serpentine channel designs, where a single winding flow channel connects the electrolyte inlet to the outlet \cite{xu2014performance,gundlapalli2019effect,xu2013numerical,ke2014flow,ke2015simple,ke2018redox,ke2017mathematical}, and interdigitated channel designs, composed of a series of interlocking channels spanning the electrode \cite{gerhardt2018effect,yin2014coupled,tsushima2020modeling,yin2019three,gundlapalli2020performance}. 
A key difference between the two is that interdigitated flow fields force all of the electrolyte through the electrode while serpentine designs allow fluid bypass. 
Thus, for the latter, only a fraction of the flowing reactant penetrates the electrode \cite{ke2018rechargeable}. 
Gerhardt et al. \cite{gerhardt2018effect} investigated the interdigitated flow field both computationally and experimentally. 
They showed how losses due to fluid pumping and the electrical overpotential change with the widths of the fluid channels and solid lands. 
While the former is generally reduced with larger fluid channel width, the latter is generally reduced with larger solid land width. 
Furthermore, for their specific system, they identified channel and land widths that lead to a minimum in overall power loss, though these dimensions changed as the operating conditions were varied. 
While this, and related, work have elucidated many of the associated transport mechanisms, there remains much debate about choosing an optimal flow field geometry \cite{Ke:2018dfa, Esan:2020jb}.
This difficult question is exacerbated by the vast design space possible for each architecture, including the recognition that the optimal flow field could change depending on the chemistry employed, the operating conditions, and the scale, etc.

Indeed, recent work is now beginning to focus on more widely exploring this design space by studying flow fields which deviate from standard 2D (i.e., a geometry that only varies in two dimensions) interdigitated and serpentine configurations \cite{LISBOA2017322, ZENG2019435, GUNDLAPALLI2021229409, akuzum2019obstructed}. 
This includes introducing hierarchy \cite{ZENG2019435}, splitting and reconfiguration \cite{GUNDLAPALLI2021229409}, and also the inclusion of 3D features such as ramps and obstructions \cite{akuzum2019obstructed}. 
A central focus of these efforts is to increase performance at larger scales where standard flow field designs may be less effective \cite{Ke:2018dfa, ZENG2019435, akuzum2019obstructed}. 

Despite their novelty, these architectures remain predominantly two dimensional, driven in large part by manufacturing constraints. 
However, recent advances in additive and advanced manufacturing offer viable techniques for fabricating complex 3D parts for electrochemical systems including flow fields \cite{ambrosi20203d, chisholm20143d, hudkins2016rapid, yang2018fully, yang2018bipolar, bui20203d}. 
Additively manufactured, porous electrodes which include 3D features have been demonstrated to lead to improved mass transfer at higher flow rates \cite{beck2021inertially}. 
There is also evidence that introducing three-dimensional features into the flow field itself can provide high performance at industrially relevant scales, e.g., as demonstrated by the design of the flow field used in the Toyota Mirai, a fuel cell vehicle \cite{yoshida2015toyota}. 

However, even as additive and advanced manufacturing technologies improve, the community lacks guidance on what 3D features to consider for these 3D flow fields, i.e., flow fields with significant geometry variation in all three dimensions. 
This difficulty can be attributed to the complex nature of the underlying physical phenomena that couples together simultaneous fluid transport, mass transfer, and electrostatics. 
While it would be possible to iterate on 3D designs using a combination of intuition and simulation, such a process would be tedious, computationally expensive, and slow. 
Instead, we propose to use topology optimization \cite{sigmund2013topology} to computationally guide flow field design. 
This method has seen significant advancements in recent years, and researchers have used topology optimization to aid in the design of heat exchangers \cite{dilgen2018density,kobayashi2019freeform, hoghoj2020topology,feppon2021body,salazar2021two,de2021three}, structural supports \cite{bendsoe1989optimal,bruns2001topology}, and microfluidic devices \cite{gersborg2005topology,borrvall2003topology}. 
Additionally, topology optimization has been applied to the design of two-dimensional flow fields for both flow batteries \cite{yaji2018topology, chen2019computational} and fuel cells \cite{behrou2019topology}. 
More recently, Beck et al. \cite{beck2021computational} demonstrated the application of three-dimensional optimization to electrochemical transport, by designing an optimized and homogenized variable porosity electrode.
Here, we leverage the ideas of that work, and develop a topology optimization method to design 3D flow fields.

In this work, we aim to use topology optimization, in tandem with a high-resolution electrochemical model of a redox flow battery, to design 3D flow fields. 
We specifically focus on the well-studied vanadium redox flow battery \cite{li2011stable,zheng2014development,davies2018high,wang2014analysis,shah2008dynamic,ma2011three}, the most widely commercialized redox flow battery \cite{kear2012development}. 
We seek to improve upon the popular interdigitated design, and in particular, we aim to showcase the pertinent three-dimensional features that arise. 
We perform the optimization to reduce the sum of electrical and fluid flow pumping power losses, and we identify how the qualitative design features of the optimized design change as the operating conditions of the flow battery change. 
We quantify the power efficiencies of the optimized designs, showcase how different optimal designs appear depending on the initial design provided to the optimization algorithm, and discuss how the optimized designs more evenly distribute reactant to the electrode. 
Since flow fields, and flow manifolds in general, are necessary for any engineered system requiring fluid and mass distribution, we also provide suggestions for how our work can be extended in the future.

\section{Methods\label{sec:methods}}

\subsection*{Problem overview}

\begin{figure}[t]
\centering
\includegraphics[width=6.4in]{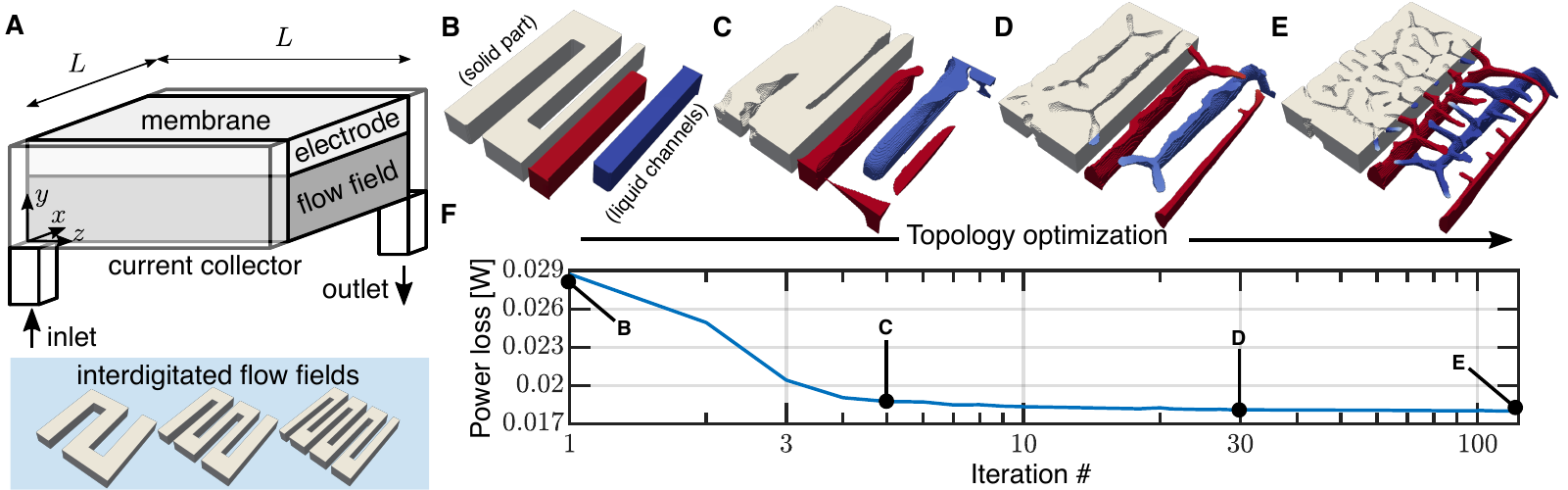}
\caption{(A) Schematic of the negative half-cell of the flow battery setup. 
A porous electrode is sandwiched between the membrane and flow field plate. 
Shown below are traditional interdigitated flow fields with various land and channel widths. 
In this work, we apply topology optimization and electrochemical modeling to further improve these designs. 
(B-E) For an example case, various iterations of the flow field designs are shown during the optimization process, corresponding to the points indicated on the plot of power loss as a function of optimization iteration shown in panel (F). 
To aid in visualization, in panels (B-E), half of the solid flow field is shown, and the negative of this, i.e. the fluid channels that form, is shown for the other half.} 
\label{fi:schematic}
\end{figure}

We consider the negative half-cell of a vanadium redox flow battery as shown in figure \ref{fi:schematic}(A). 
Liquid electrolyte, comprised of a solution of $V^{2+}$ and $V^{3+}$ at a concentration $C_\mathrm{inlet} = 1\mathrm{M}$ in 1M sulfuric acid flows in, which is then guided by the flow field towards the porous carbon-felt electrode, where the reaction $V^{2+} \rightarrow V^{3+} + e^- $ occurs on the surfaces of electrode fibers. 
The spent fluid then flows from the outlet. 
Inspired by the popularity and success of the interdigitated flow field design (examples of which are shown in the bottom of figure \ref{fi:schematic}(A)), we utilize a combination of a continuum model of the flow battery half-cell following porous electrode theory \cite{beck2021computational,newman1975porous} -- incorporating fluid mechanics, mass transfer, and electrostatics -- and topology optimization to improve upon this design. 
An example of the optimization process used in this work is shown in figures \ref{fi:schematic}(B-E).
Specifically, the flow field designs at various iterations in the optimization process are shown, corresponding to points indicated on the plot of power loss plotted against iteration number in figure \ref{fi:schematic}(F). 
To aid in visualization, in figures \ref{fi:schematic}(B-E), half of the solid flow field is shown, and the negative of the solid part, i.e. the fluid channels within the flow field, are shown for the other half.
In this work, we focus on a half-cell with experimentally relevant dimensions: the square porous carbon electrode has side length of $L = 2$ cm, and thus an area of $A=4$ $\mathrm{cm^2}$, and is $500$ $\mathrm{\mu m}$ thick, and the flow field is 3 mm thick.

We aim to design the flow field to minimize the excess power losses $P$ that occur for a specified discharge current $I$ and electrolyte flow rate $Q$. 
We consider in this work current densities $I/A=\{50,100,150\}$ $\mathrm{mA/cm^2}$ and flow rates $Q=\{2.4, 12, 24, 36\}$ $\mathrm{mL/min}$ \cite{shah2008dynamic, li2011stable,gerhardt2018effect,davies2018high,ma2011three}. 
The total power loss is defined as
\begin{equation}
P =\int\limits_{\mathrm{membrane}}\eta  \frac{I}{A}\,\rmd \mathbf{x} +\frac{1}{\Psi_\mathrm{pump}}\int\limits_{\mathrm{inlet}}p\bar{u}_\mathrm{inlet}\,\rmd \mathbf{x}, \label{eq:power}
\end{equation}
where $\eta$ is the half-cell overpotential, $I/A$ is the discharge current density, $\Psi_\mathrm{pump}$ is the pump efficiency, which is assumed to be $0.5$ in this work, $p$ is the fluid pressure, and $\bar{u}_\mathrm{inlet}=Q/A_\mathrm{inlet}$ is the specified averaged inlet fluid velocity. 
The first term represents the electrical power loss due to the half-cell overpotential, while the second term represents the fluid pumping power loss. 
As highlighted by Gerhardt et al. \cite{gerhardt2018effect}, a flow field design with more solid present will typically have a reduced Ohmic electrical power loss due to the high conductivity of the solid, and a flow field design with more void will typically have a reduced fluid pumping power loss due to the reduced hydraulic resistance -- this paradox highlights why optimization of this component is nontrivial. 
Equivalently, our goal is to maximize the overall power efficiency of the half-cell, which is defined as
\begin{equation}
\Xi = 1-\frac{P}{IU_0},\label{eq:powereff}
\end{equation}
where $U_0$ is the equilibrium half-cell potential. 
As defined, $\Xi=1$ for a perfectly efficient cell and can be $<0$ for a very poorly designed cell.

\subsection*{Continuum simulation}

The equations governing the continuum electrochemical model are as follows. 
The fluid velocity $\mathbf{u}$ is described by the steady incompressible Navier-Stokes equation with a Darcy drag,
\begin{eqnarray}
\rho\mathbf{u}\cdot\nabla\mathbf{u} &= - \mu\alpha^{-1}\mathbf{u} -\nabla p + \mu\nabla^2\mathbf{u},\label{eq:NS_mom}\\
\nabla\cdot\mathbf{u} &= 0,\label{eq:NS_cont}
\end{eqnarray}
where $\rho$ is the fluid density, $\mu$ is the viscosity, $p$ is the pressure, and $\alpha$ is the local permeability of the material. 
As described below, the topology optimization technique models the presence of solid and liquid material by smoothly varying the parameters using a position dependent design variable, $\gamma(\mathbf{x})$. 
In the solid, $\alpha^{-1}$ takes on a large value to prevent fluid flow, in the liquid regions, $\alpha^{-1}=0$, and in the fixed porous region representing the electrode, $\alpha = \alpha_E$. 
This parameter thus controls the location of solid and liquid within the flow field and approximates the no-slip condition for the fluid velocity. 
At the inlet, we specify the velocity to be $\mathbf{u} = -(Q/A_\mathrm{inlet})\mathbf{n}$, and the fluid exits to zero pressure; no-slip is assumed at all other boundaries. 

The mass transfer of the species concentration $C_i$, where $i\in\{V^{2+},V^{3+}\}$, is described by the steady advection-diffusion-reaction equation, given by,
\begin{equation}
\mathbf{u}\cdot\nabla C_i  = \nabla\cdot (D_i\nabla C_i) + aj_{n,i}\mathbf{1}_{\mathbf{x}\in\mathrm{electrode}},\label{eq:mass}
\end{equation}
where $D_i$ is the design variable dependent diffusivity, $a$ is the exposed surface area per volume of the porous electrode, and $\mathbf{1}_{\mathbf{x}\in\mathrm{electrode}}$ is an indicator function that returns $1$ in the electrode and $0$ elsewhere. 
The reaction $j_{n,i}$ within the homogenized electrode is given by $j_{n,i}=k_\mathrm{m}(C_i^\mathrm{s} - C_i)$, where $k_\mathrm{m}$ is a mass transfer coefficient and $C_i^\mathrm{s}$ is the concentration at the surface of the fibers of the electrode. 
At all boundaries, the no-flux condition is applied for $C_i$.

The electrostatics of the solid and liquid are described by the following Poisson equations, 
\begin{eqnarray}
\nabla\cdot\left(-\sigma\nabla \Phi_1\right) &= -ai_n\mathbf{1}_{\mathbf{x}\in\mathrm{electrode}}, \label{eq:elec1}\\
\nabla\cdot\left(-\kappa\nabla \Phi_2\right) &= ai_n\mathbf{1}_{\mathbf{x}\in\mathrm{electrode}}, \label{eq:elec2}
\end{eqnarray}
where $\Phi_1$ and $\Phi_2$ are electric potentials of the liquid and solid, respectively, and $\sigma$ and $\kappa$ are the design variable dependent conductivities of the liquid and solid, respectively. 
The exchange current $i_n$ is given by the Butler-Volmer equation,
\begin{equation}
i_n = \frac{i_0}{C^\mathrm{ref}}\left(C_{V^{2+}}^s e^{\beta\Delta\Phi} - C_{V^{3+}}^s e^{-\beta\Delta\Phi}\right),\label{eq:BV}
\end{equation}
where $C^\mathrm{ref}=0.001\mathrm{M}$ is an arbitrary reference concentration to maintain dimensional consistency, $\Delta\Phi$ is the local overpotential, $\beta=0.5F/RT$, where $F$ is Faraday's constant, $R$ is the ideal gas constant, $T$ is the temperature, and a charge transfer coefficient of $0.5$ has been assumed.
The current $i_n$ is proportional to the mass transfer reaction term as $i_n = Fj_{n,V^{2+}} = - Fj_{n,V^{3+}}$.
A current is discharged through the membrane via the liquid, so that $-\kappa\nabla\Phi_2\cdot\mathbf{n} = I/A$ at the membrane, and the current collector is attached to the solid part of the flow field so that $\Phi_1 = 0$ at the current collector; all other boundaries are no-flux boundaries for the potential.

\subsection*{Topology optimization}

We follow the optimization method described by Beck et al. \cite{beck2021computational}, with some important differences. 
In contrast to \cite{beck2021computational}, who performed optimization of a porous electrode via a homogenized model, we utilize topology optimization to generate flow field designs with well-defined solid and liquid boundaries separated by a thin, smoothly varying interface \cite{sigmund2013topology}. 
Our goal is to solve,
\begin{eqnarray}
&\min_{\gamma \in [0,1]} P\\
\mathrm{s.t.}\, &\mathbf{R}_{\{\mathbf{u},p,\Phi_1,\Phi_2,C_i\}} = \mathbf{0},
\end{eqnarray}
where $\mathbf{R}_{\{\mathbf{u},p,\Phi_1,\Phi_2,C_i\}}=\mathbf{0}$ are the forward partial differential equations (PDEs) given in equations (\ref{eq:NS_mom})-(\ref{eq:elec2}), and $\gamma(\mathbf{x})$ is the design variable within the flow field that is equal to $1$ where the flow field is pure liquid and equal to $0$ where the flow field is pure solid. 

Within the flow field, the design variable $\gamma(\mathbf{x})$ is used to smoothly interpolate the material properties from that of pure solid to that of pure liquid. 
Thus, the permeability, diffusivities, and conductivities in equations  (\ref{eq:NS_mom})-(\ref{eq:elec2}) are all functions of the design variable.
To ensure that optimized designs are ``black-white,'' i.e. comprised of pure solid and pure liquid with minimal material in between, the specific functional forms of the interpolations are chosen carefully (e.g. see \cite{dilgen2018density}) and attain the appropriate values at the end points, $\gamma=0$ and $\gamma=1$. 
These functions are described in the Supplementary Information. 

To perform the optimization, the continuous adjoint approach is employed \cite{othmer2008continuous,beck2021computational}.
Briefly, the Lagrangian of the problem is
\begin{equation}
L = P+\int\mathbf{S}\cdot\mathbf{R}\,\rmd\mathbf{x}.
\end{equation}
The adjoint PDEs $\mathbf{S}$ corresponding to each PDE given in equations (\ref{eq:NS_mom})-(\ref{eq:elec2}) are derived analytically by setting the variation of $L$ with respect to each dependent variable to 0, i.e. $\delta_{f_i}L=0$. 
From this, the sensitivity is then determined by calculating $\delta L=\delta_\gamma L$. 
These equations are solved in OpenFOAM, and the sensitivities are used within the Method of Moving Asymptotes to update the values of $\gamma$ in each computational cell \cite{Svandberg:1987xi}. 
For the simulations in this work, we use hexahedral meshes with $4.3 - 12.6$M cells to solve the forward and adjoint PDEs.
The Helmholtz filter with a filter radius $r$ of $2\times$ the mesh size used in the flow field, $r=167$ $\mu\mathrm{m}$, is used to regularize the solution, and a Heaviside projection is used to push the solution to a black-white design \cite{lazarov2011filters}. 
This process is described in more detail in the Supplementary Information.

\subsection*{Estimation of electrochemical parameters}

Estimated parameters and correlations used in our electrochemical simulation are shown in table \ref{ta:params}. 
Notably, the Bruggeman correlation \cite{shah2008dynamic} is used to describe properties within the porous electrode. 
Specifically, in the electrode,
\begin{eqnarray}
D &= D_0\epsilon^{3/2},\\
\kappa &= \kappa_0\epsilon^{3/2},\\
\sigma &= \sigma_0(1-\epsilon)^{3/2},
\end{eqnarray}
and in the flow field, interpolations relating the design variable to solid and liquid properties are described in the Supplementary Information.

\begin{table}
\caption{\label{ta:params}Parameters used for electrochemical model} 
\begin{indented}
\lineup
\item[]\begin{tabular}{@{}*{7}{l}}
\br                              
Description & Parameter & Value & Units & Ref. \cr
\mr
Solid conductivity 			& $\kappa_0$ 			 & $10^4$				 & $\mathrm{S/m}$ 		& Est.\cr
Electrode porosity 			& $\epsilon$ 			 & $0.68$ 				 & - 			  		& \cite{shah2008dynamic}\cr
Electrode permeability 		& $\alpha_E$ 		     & $5.53\cdot 10^{-11}$  & $\mathrm{m^2}$ 		&\cite{shah2008dynamic}\cr
Electrode area/volume 		& $a$ 					 & $8\cdot10^4$ 		 & $\mathrm{m^{-1}}$	& Est. \cr
Electrolyte conductivity 	& $\sigma_0$ 			 & $40$ 				 & $\mathrm{S/m}$ 		& Est.\cr
Electrolyte viscosity   	& $\mu$ 			     & $8.9\cdot10^{-4}$ 	 & $\mathrm{Pa\cdot s}$ & Est.\cr
Electrolyte density	    	& $\rho$ 			     & $10^{3}$ 	 	     & $\mathrm{kg/m^{3}}$ 	& Est.\cr
Diffusivity 				& $D_0$ 				 & $2.4\cdot 10^{-10}$   & $\mathrm{m^2/s}$ 	& \cite{yamamura2005electron}\cr
Mass transfer coeff. 		& $k_m$ 		 		 &  $4.4\cdot10^{-4}|\mathbf{u}|^{0.4}$ & $\mathrm{m/s}$ & \cite{schmal1986mass}\cr
Exchange current 			& $i_0$ 				 & $0.016$ 				 & $\mathrm{A/m^2}$     &\cite{shah2008dynamic}\cr
Equilibrium potential 		& $U_0$				     & $-0.25$ 				 & $\mathrm{V}$		    & \cite{li2011stable} \cr
\br
\end{tabular}
\end{indented}
\end{table}

\section{Results and discussion}

\subsection*{Optimized flow field designs}

\begin{figure*}[t]
\centering
\includegraphics[width=13.8cm]{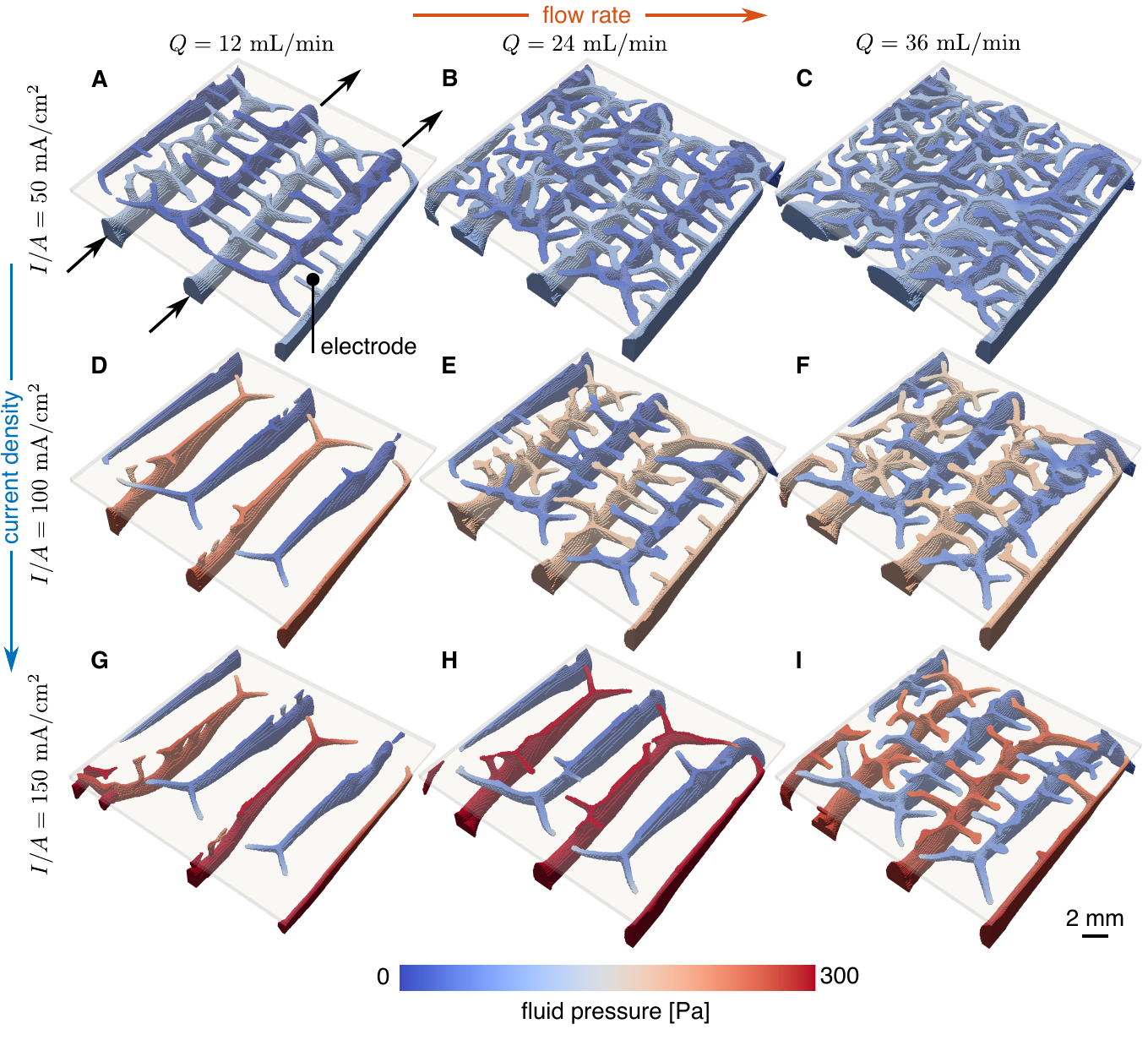}
\caption{The fluid fraction of the optimized flow field, colored by fluid pressure, is shown for various flow rates $Q$ and current densities $I/A$. 
The porous electrode is shown as a translucent sheet. 
Fluid flows in through the higher pressure channels and up into the porous electrode, reacts, and flows back out through the lower pressure channels. 
An initial channel and land width of $2$ $\mathrm{mm}$ is used. 
(A-C) $I/A=50$ $\mathrm{mA/cm^2}$, (D-F) $I/A=100$ $\mathrm{mA/cm^2}$, (G-I) $I/A=150$  $\mathrm{mA/cm^2}$.} 
\label{fi:ff}
\end{figure*}

\begin{figure*}[t]
\centering
\includegraphics[width=3.25in]{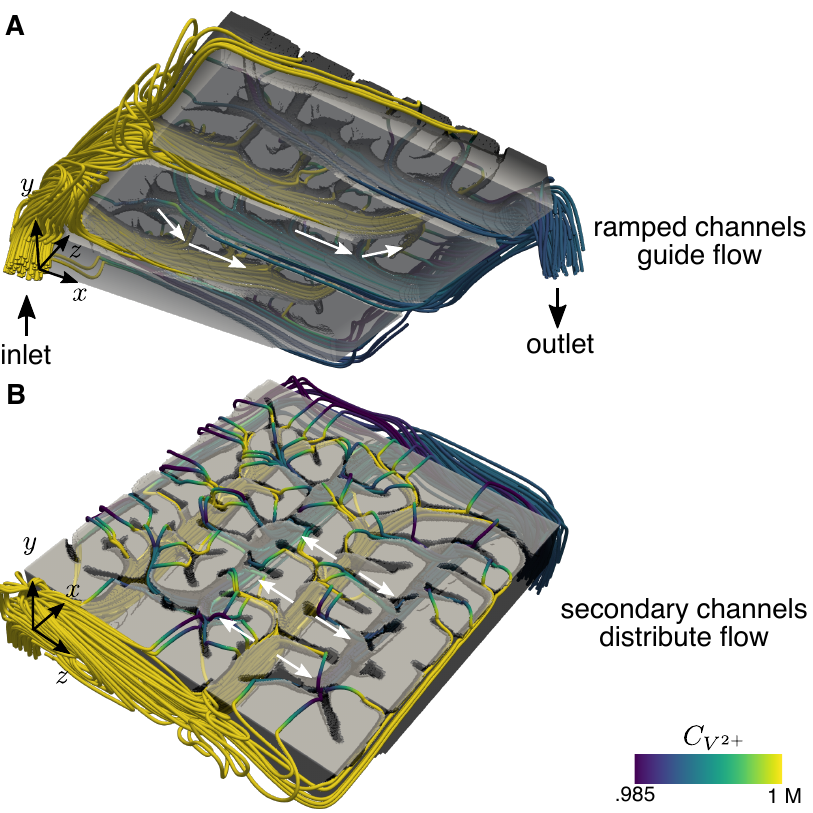}
\caption{Visualization of the fluid streamlines, colored by reactant concentration, through the solid flow field for the example case where $I/A=100$ $\mathrm{mA/cm^2}$ and $Q  = 24$ $\mathrm{mL/min}$. 
Panel (A) shows a view of the bottom half of the flow field that interfaces with the current collector. 
Ramps appear at the ends of the channels to guide flow into and out of the electrode. 
Panel (B) shows a view of the top half of the flow field that interfaces with the membrane. 
Small secondary channels appear to evenly spread reactant across the electrode. 
In both panels, the solid flow field is slightly transparent and shaded to aid in visualization. 
}
\label{fi:flow}
\end{figure*}

In figure \ref{fi:ff}, we show the liquid fraction of optimized flow fields for a range of operating conditions where the optimization is initialized with channel and land widths of 2 mm. 
In figures S1 and S2 of the Supplementary Information, results are also shown where the optimization is initialized with channel and land widths of 4 and 1.5 mm, respectively.  
The actual solid part is thus the negative of what is shown in the figure. 
The porous electrode is represented as a translucent sheet above the visible flow structures.
As the operating current and flow rate change, the dominant physical phenomena change. 
This, in turn, changes the ratio of electrical power losses to pressure power losses, leading to qualitatively different optimal flow field designs. 
When the flow rate is low, the optimized fluid channels are thin and the flow field is comprised mostly of solid material, suggesting that an interdigitated design with thin channels would perform well.
At even lower flow rates, we observe no clear three-dimensional structures that form in the nearly fully-solid flow field, and these designs are shown in figure S3 of the Supplementary Information.
In contrast, as the flow rate increases and more reactant is supplied to the electrode, smaller secondary channels emerge from the primary fluid channels beneath -- resembling a tree-root structure -- and guide fluid up into the electrode. 
Additionally, the larger primary fluid distribution channels grow larger, presumably to avoid large fluid resistances and increased flow power losses. 

The observed features of the optimized designs are fully three-dimensional, providing evidence that there is room for improvement in more popular two-dimensional designs. 
Remarkably, the optimization algorithm  automatically identified features which have previously been devised through experimentation. 
At flow rates of $Q=$12 mL/min and 24 mL/min, the optimized designs show a ramped channel design as in the work of Akuzum et al. \cite{akuzum2019obstructed}, helping to guide both unreacted reactant up into the electrode and used reactant down out of the electrode. 
The optimized design, however, does further taper the channel in addition to ramping. 
Additionally, at high flow rates and currents, the optimization algorithm not only re-identified the use of hierarchy, but it further suggests that the density of branching should increase at lower operating currents and higher flow rates \cite{ZENG2019435}. 
The optimization procedure not only suggests interesting design candidates, but because it uses a model and simulation for the system behavior to determine the geometries, it also provides strategies to most effectively employ the 3D features.
To provide further intuition on the design features that form and how fluid flows through the solid part, in figure \ref{fi:flow}, we show a visualization of the fluid streamlines through the flow field for the example case where $I/A=100$ $\mathrm{mA/cm^2}$ and $Q  = 24$ $\mathrm{mL/min}$. 
These streamlines are colored by reactant concentration $C_{V^{2+}}$ to highlight the depletion of concentration as fluid leaves the flow field.
In figure \ref{fi:flow}(A), we observe the previously described ramps that appear at the ends of the primary fluid distribution channels, and in figure \ref{fi:flow}(B), we observe the smaller secondary channels that form at the flow field-electrode interface that work to evenly distribute fluid across the electrode. 

In our optimization procedure, it is important to note that the length scale of the smallest feature size can, in principle, be controlled to meet minimum feature size constraints imposed by a specific additive manufacturing technique. 
As described in \S \ref{sec:methods}, in these optimization simulations, a Helmholtz filter \cite{lazarov2011filters} is used to regularize the flow field design, which as a result sets the minimum feature size that can arise here. 
In this work, we have chosen a filter radius that results in a minimum feature size of approximately $0.5$ mm, however, this filter radius could be modified.
Current additive manufacturing techniques have resolutions that range from sub-micron to millimeter scale \cite{ambrosi20203d}.

\subsection*{Power losses and efficiencies}

\begin{figure*}[t]
\centering
\includegraphics[width=6.7in]{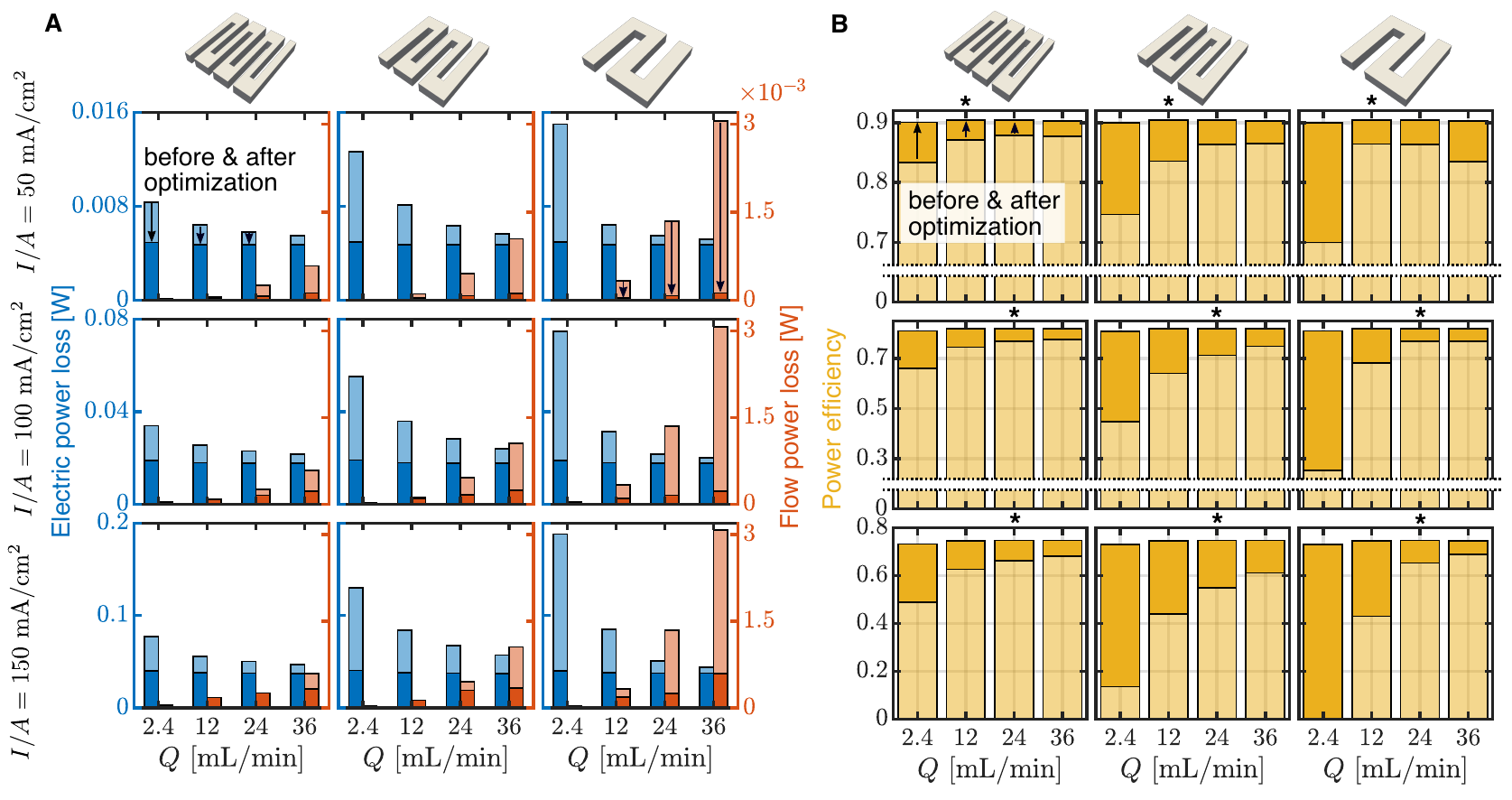}
\caption{(A) Electric and flow power losses of the unoptimized and optimized flow fields for various flow rates $Q$, current densities $I/A$, and initial conditions. 
Current densities $I/A=\{50,100,150\}$ $\mathrm{mA/cm^2}$, flow rates $Q=\{2.4, 12, 24, 36\}$ $\mathrm{mL/min}$, and flow fields initialized with land and channel widths of $\{1.5, 2, 4\}$ $\mathrm{mm}$ are considered. 
(B) Corresponding power efficiencies $\Xi=1-P/(IU_0)$. 
Asterisks indicate the flow rate yielding the highest power efficiency. 
In the individual bars, the light portion of the bar shows the value before the optimization procedure and the dark portion the value after, as indicated by the arrows.}
\label{fi:powerloss}
\end{figure*}

The qualitative design features observed in figure \ref{fi:ff} are now quantitatively supported. 
As implied by the variety of designs shown in figure \ref{fi:ff}, the optimal flow field design is achieved by balancing the electrical and flow power losses. 
This balance is similar to what is observed in the optimized homogenized porous electrodes studied by Beck et al. \cite{beck2021computational}, where lower operating flow rates lead to less porous optimized structures and vice versa. 
The associated electric and flow power losses of the optimized designs are shown in figure \ref{fi:powerloss}(A), along with the power losses obtained with the traditional, unoptimized interdigitated flow field. 
The power efficiencies $\Xi$, calculated from equation (\ref{eq:powereff}), corresponding to these power losses are shown in figure \ref{fi:powerloss}(B). In all cases, the optimized designs lead to greater power efficiency than the interdigitated designs.

\begin{figure*}[t]
\centering
\includegraphics[width=11.4cm]{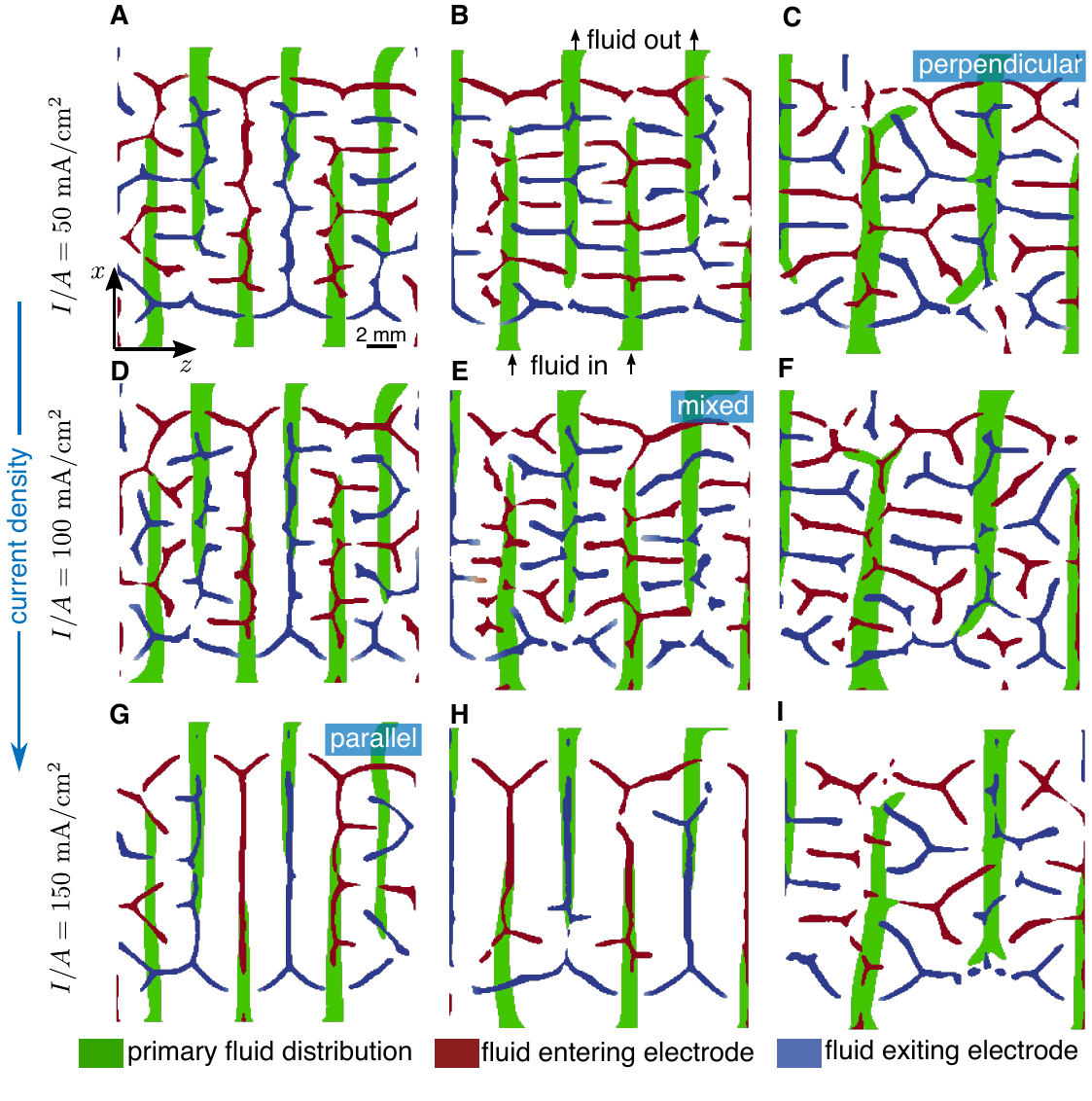}
\caption{Optimized fluid distribution channels. 
Slices of the liquid fraction of the flow field are shown at $y=1$ mm (green) and close to the flow field-electrode interface ($y=2.9$ mm). 
Current densities $I/A=\{50,100,150\}$ $\mathrm{mA/cm^2}$ are considered, and flow rates $Q$ are chosen to correspond to the optimal power efficiency as shown in figure \ref{fi:powerloss}(B). 
The different panels show different starting interdigitated designs; specifically the initial land and channels widths are (A,D,G) 1.5 mm (B,E,H) 2 mm and (C,F,I) 4mm. 
As the initial condition changes, the optimization procedure leads to qualitatively different designs: secondary channels that are largely perpendicular to the primary flow distribution channels appear when the initial channel and land widths are thick, whereas they are parallel when the  initial channel and land widths are thin.}
\label{fi:3D}
\end{figure*}

At a particular current, traditional flow field designs yield power losses where the electric losses are maximized when the supplied flow rate is minimized, and the flow losses are maximized when the supplied flow rate is maximized. 
Thus, the optimization prioritizes reduction of the appropriate power loss contribution based on the operating conditions. 
This is evident in figure \ref{fi:powerloss}(A) from the large reduction in the electric power loss for designs generated for low flow rates, and, conversely, in the large reduction in flow power loss for designs generated for high flow rates. 
For the particular system studied here, the electrical power losses typically dominate, with the flow power losses typically contributing $<10\%$, due to the fact that the dominant source of flow resistance is only associated with the thin porous electrode. 
For larger area electrodes, the flow losses would be larger, and this optimization process would appropriately reprioritize the loss reduction. 
Since the specific choice of channel and land widths of traditional interdigitated flow fields are known to affect performance, we also consider simulations initialized with channel and land widths of 1.5 and 4 mm, and observe the same trends for each initial condition. 

For a given current, the power efficiency of the optimized design reaches approximately the same value across all of the flow rates and initial conditions tested. 
For example, when $I/A=100$ $\mathrm{mA/cm^2}$, the optimized power efficiency $\Xi$ is always $\approx 0.82$. 
In contrast, at the same current density, the standard interdigitated designs have efficiency $\Xi\approx 0.25$, when $Q=2.4$ $\mathrm{mL/min}$ and the channel and land widths are 4mm, while $\Xi\approx 0.78$, when $Q=36$ $\mathrm{mL/min}$ and the channel and land widths are 1.5 mm. 
The improvement from optimization at this current density thus ranges from $5-228\%$. 
This demonstrates that topology optimization can be an effective tool to identify a high performance geometry regardless of the desired operating conditions. 
Alternatively, a single standard unoptimized design shows strong variations in performance across operating conditions. 
While it is certainly possible to use a traditional interdigitated design to yield high performance, a significant amount of testing is required to find the optimal fluid channel and solid land widths for the specific system used; this process needs to be repeated whenever these conditions change. 
For example, Gerhardt et al. \cite{gerhardt2018effect} find a channel and land width that yields a maximum pumping-corrected voltage efficiency, a metric similar to our definition of $\Xi$, of $\approx 80\%$ for their ferrocyanide-ferricyanide flow battery system operating at $250$ $\mathrm{mA/cm^2}$, but also discuss how the optimal dimensions change as the operating conditions change. 
A key takeaway from the present work is that this iterative process can be automated through the use of topology optimization methods such as the one presented here.

\subsection*{Analysis of 3D flow field design features}

All of the flow field designs presented in figure \ref{fi:ff} have been obtained by conducting simulations with the same initial condition: an interdigitated flow field with channel and land width of $2$ $\mathrm{mm}$. 
However, as evidenced by the power efficiencies presented in figure \ref{fi:powerloss}(B), the optimization scheme is able to yield a design with similar performance, regardless of the initial channel and land widths chosen, even though the ultimate design depends on the specific initial conditions. 
To further understand how the optimization procedure is selecting design features, in figure \ref{fi:3D}, we show slices of the fluid fraction of the optimized flow fields from figures \ref{fi:ff}, S1, and S2, where the flow rate $Q$ is selected based on the optimal power efficiency as observed in figure \ref{fi:powerloss}(B).
In doing so, we highlight how the optimized three-dimensional structures change for various currents and starting initial conditions. 
In green, the primary underlying fluid distribution channels are shown by plotting the fluid fraction along a slice within the flow field ($y=1$ mm), and in blue and red, the secondary fluid channels are shown by plotting the fluid fraction along a slice near the flow field-electrode interface ($y=2.9$ mm). 
The fresh reactant-laden fluid thus enters the flow field through the green channels on the bottom, enters the electrode in the out-of-page $y$ direction through the red high pressure regions, exits the electrode through the blue low pressure regions, and out of the device through the green channels on the top.

The structure of the secondary channels that form provide clues for what constitutes an optimal design. 
In figure \ref{fi:3D}(C), where the current is relatively low and the initial interdigitated design has relatively thick channel and land widths, secondary channels appear perpendicular to the primary flow channels. 
In contrast, in figure \ref{fi:3D}(G), where the opposite is true, secondary channels appear parallel and coincident to the primary flow channels. 
In other cases, a mixed regime appears, where both parallel and perpendicular secondary channels appear. 
This provides evidence that the initial channel and land widths in figure \ref{fi:3D}(C) are too thick for the system, and the optimization algorithm works to create secondary channels to more evenly distribute reactant. 
Similarly, this shows that the initial channel and land widths in figure \ref{fi:3D}(G) are close to optimal, and the interdigitated design under these operating conditions already performs relatively well. 
In figure \ref{fi:3D}(A), we observe a design resembling that of figure \ref{fi:3D}(G), but with small perpendicular channels appearing: since the operating current in this case is lower, there is less need for the optimization to prioritize electrical contact with the flow field.

\begin{figure*}[t]
\centering
\includegraphics[width=11.4cm]{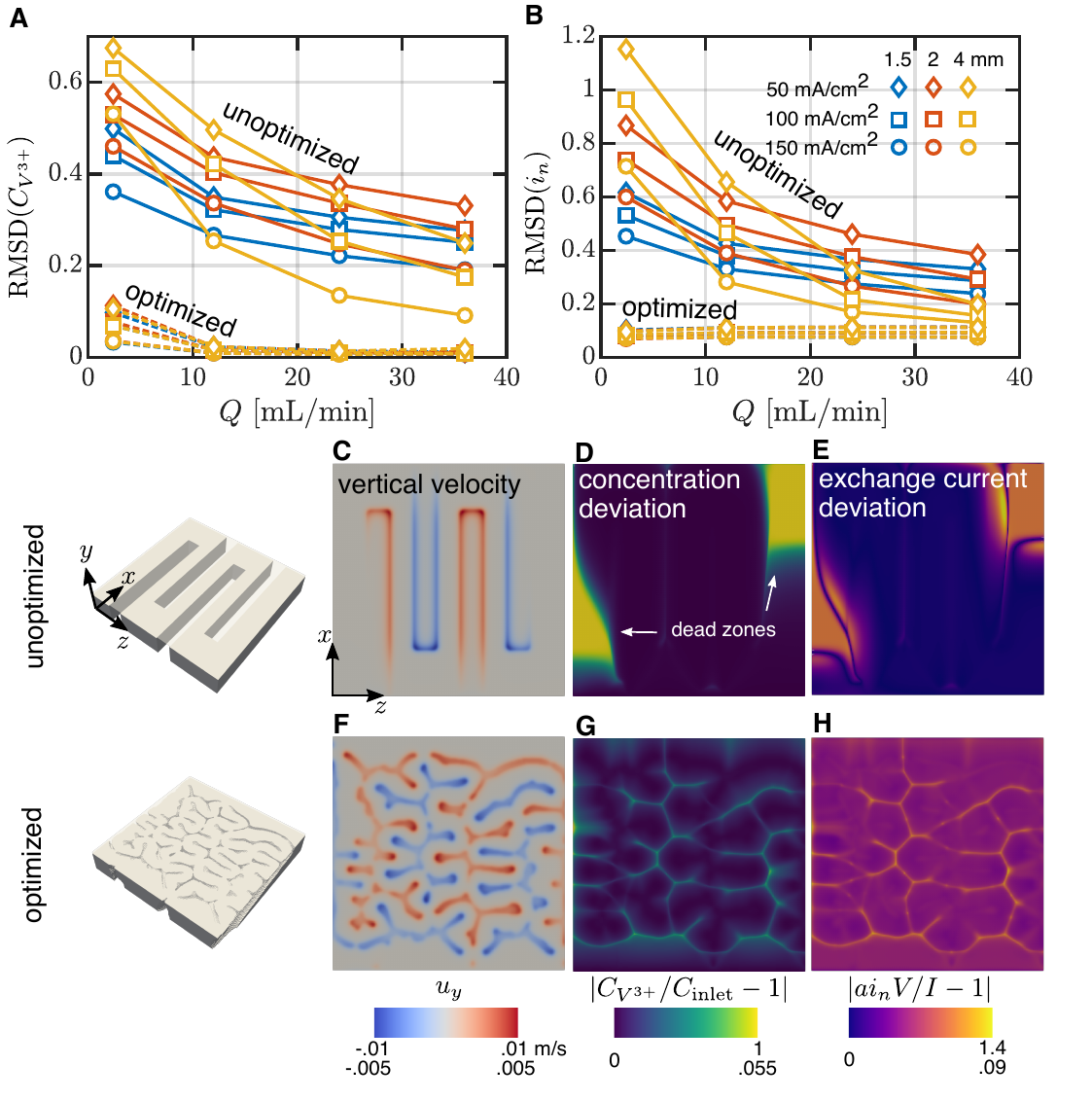}
\caption{Dead zones of the electrode. 
Panels (A) and (B) show the root-mean-squared deviation of the species concentration and the exchange current, respectively, within the electrode for both the unoptimized and optimized flow fields. 
Panels (C-H) show the local vertical velocity, concentration polarization and exchange current polarization at the center of the porous electrode for the specific case of $I/A=100$  $\mathrm{mA/cm^2}$, $Q=24$  $\mathrm{mL/min}$, and an initial channel and land width of $2$ $\mathrm{mm}$. 
The images to the left show the solid fraction of the corresponding flow field. 
The optimized flow field more evenly distributes reactant, leading to more uniform reaction and a reduction in the extent of dead zones in the electrode. 
In the color legend, numbers on the top correspond to panels (C-E) and numbers on the bottom correspond to panels (F-H).}
\label{fi:RMS}
\end{figure*}

To further understand and quantify how evenly the reactant is distributed, we consider the dimensionless root-mean-squared-deviation of the oxidant concentration within the electrode,
\begin{equation}
\mathrm{RMSD}(C_{V^{3+}}) = \sqrt{\left< \left(\frac{C_{V^{3+}}}{C_\mathrm{inlet}} - 1\right)^2 \right>},
\end{equation}
where $\left<\cdot\right>$ denotes an average over the volume of the porous electrode. 
This quantity thus gives a measure of the concentration polarization within the electrode. 
Similarly, we also consider the dimensionless root-mean-squared-deviation of the exchange current $i_n$ within the electrode,
\begin{equation}
\mathrm{RMSD}(i_n) = \sqrt{\left< \left(\frac{ai_nV}{I} - 1\right)^2 \right>},
\end{equation}
where $a$ is the exposed surface area per volume of the electrode, $V$ is the electrode volume, and $i_n$ is the exchange current  given by the Butler-Volmer relation in equation (\ref{eq:BV}).
It is evident that the lower these quantities are, the more uniform the distribution of reactant and Faradaic reaction. 
In the limit of perfect uniformity, both approach a value of zero.
In figure \ref{fi:RMS}(A) and (B), we show both $\mathrm{RMSD}(C_{V^{3+}})$ and $\mathrm{RMSD}(i_n)$ as a function of flow rate for various currents and initial channel and land widths. 
In all cases, we observe an appreciable reduction in these quantities for the optimized designs, with the most significant improvement observed at low flow rate.

While both of these quantities give a global measure of how well the reactant is distributed within the electrode, it is also insightful to examine the local quantities. 
In figure \ref{fi:RMS}(C)-(H), we compare the vertical fluid velocity, local concentration polarization, and local exchange current deviation within a slice of the electrode between the unoptimized and optimized flow field; we consider the particular case where $I/A=100$  $\mathrm{mA/cm^2}$, $Q=24$  $\mathrm{mL/min}$, and the initial channel and land width is $2$ $\mathrm{mm}$. 
This highlights how the optimized design is able to more evenly utilize the entire electrode and avoid electrode dead zones. 
In figure \ref{fi:RMS}(C), we observe that there is a large area of the electrode where the fluid is essentially stagnant. 
In contrast, the optimized design is able to deliver fluid to and from the electrode nearly uniformly (figure \ref{fi:RMS}(F)), thus reducing the concentration and exchange current polarization (figure \ref{fi:RMS}(G,H)).

\section{Conclusions}

In this work, we have shown that topology optimization, coupled with a model of the fluid mechanics, mass transfer, and electrostatics describing a flow battery, can be used to improve the performance of a flow field. 
Specifically, we use topology optimization to design flow fields with full three-dimensional variation, i.e. 3D flow fields.
We have improved upon the popular two-dimensional interdigitated flow field design and identified pertinent three-dimensional features, such as ramps and hierarchy, that have previously been devised heuristically.
Additionally, we have shown that our optimized designs lead to better overall reactant and reaction distribution within the electrode.  
Notably, while using a traditional interdigitated design can certainly yield high performance, this requires tedious testing to find the optimal land and channel widths for the particular system; our method has the potential to automate that process. 
Understanding and improving fluid and mass distribution in electrochemical systems such as flow batteries is critical, and this work gives preliminary evidence that investing effort into developing advanced manufacturing techniques to fabricate such complex components may be fruitful. 
Our ongoing work includes experimentally demonstrating that the flow fields like the ones described in the present work can provide real-world benefit.

There are a number of other considerations that may need to be taken into account in an industrial setting for a specific application. 
For example, one may be interested in maximizing the single-pass conversion efficiency of the reactant, instead of the power loss, when considering a system-level analysis of a flow battery stack. 
Next, while we have focused on a particular flow battery size in this work, it would be of interest to explore how the benefits of optimization are affected as the device is scaled up.
Certainly, the ratio of electrical losses to flow losses would be expected to change with system size. 
Furthermore, it would be of interest to understand how the optimized flow fields in this work perform within a full cell, where the membrane would be expected to make a nontrivial contribution to the performance.
In a real-world application, it would also be important to understand whether or not these optimized designs would perform well in the presence of obstacles such as bubbles or unwanted debris that can clog flow channels.
Finally, while flow batteries have been subject of the present work, the distribution of reactants remains to be a key problem in a range of engineered systems, such as $\mathrm{CO_2}$ \cite{wheeler2020quantification} and water electrolyzers, fuel cells, and bioreactors. 
Given a particular system chemistry and desired quantity to be minimized, the process outlined in this work can be repeated, leading to the potential to improve mass distribution in such systems. 

\ack{
T.Y.L. thanks Drs. A. Wong, J. Davis, B. S. Jayathilake, and S. Chandrasekaran for insightful conversations about interdigitated flow fields and additive manufacturing. 
This work was performed under the auspices of the U.S. Department of Energy by the Lawrence Livermore National Laboratory under contract DE-AC52-07NA27344. 
This work was also supported by the Lawrence Livermore National Laboratory LDRD 19-SI-005 and 19-ERD-035. 
This work was performed under the auspices of a cooperative research and development agreement between the Lawrence Livermore National Laboratory, Stanford University, and TOTAL American Services, Inc. (affiliate of TOTAL SE) under agreement number TC02307.
LLNL release number: LLNL-JRNL-832021-DRAFT.
}

\section*{References}

\bibliographystyle{unsrt}
\bibliography{flowfieldopt_preprint}

\end{document}